\documentstyle[twocolumn,aps]{revtex} 
\begin{document}

\title{Condensed Matter Physics - Biology Resonance }

\author{G. Baskaran\cite{email}}

\address{Institute of Mathematical Sciences\\
C I T Campus\\
Madras 600 113, India}

\maketitle

\begin{abstract}

The field of condensed matter physics had its genesis this century
and it has had a remarkable evolution. A closer look at its growth  
reveals a hidden aim in the collective consciousness of the field
- a part of the development this century is a kind of 
warm up exercise to understand the nature of living condensed matter, 
namely the field of biology, by a growing new breed of scientists 
in the coming century. Through some examples the vitality of this 
interaction will be pointed out.

\end{abstract}

\section{Introduction} 

`Condensed Matter Physics' (CMP)  is a clever name for the study of 
any form of matter that is condensed - liquids, solids, gels,
cells, superfluid He4, quantum Hall liquid etc etc.  The folklore
is that the name `Condensed Matter Theory' was coined in Cambridge 
in the 60's in the solid state group that involved people like V. Heine 
and P.W. Anderson.  Of course, before this field was christened
it existed, on its own right as solid state physics and related
fields with very many significant developments to its credit - 
the new name gave it an added identity and perhaps a new purpose.  

Physics, a part of Natural Science, is an experimental science.
It gains its strength from experiments, observations, theorizing,  
and impact on technology and society. CMP has a special place in physics
because of its closeness to a multitude of feasible and often novel
experiments.  This feasibility is intimately tied with the wealth of
matter and associated phenomena around as well as the 
development in the field of material science and in turn technology - 
both low and high tech.  Elegant concepts from quantum physics, 
statistical mechanics, mathematics are combined alive so that it 
continues to produce surprises and new phenomena and new concepts.  
This field is also a source for innovative new experimental methods
that has extended human `senses' to atomic scale - modern x-ray 
crystallography, NMR, neutron scattering, spectroscopy, scanning 
tunneling microscope and so on.

The aim of the present article is to provide  a point of view 
that this field has grown, partly  with an aim to address deeper 
issues in the field of living condensed matter namely biology;
and a century of efforts is really a warm up exercise towards
this difficult goal.  The point of view I am providing is perhaps
obvious  - my main message is that a true resonance between
the two fields is something that is natural and so likely to 
happen or has already begun.

\section{Nature of Condensed Matter Physics}

CMP is diverse and complex.  It addresses issues such as why
silicon has a diamond like structure using quantum mechanical 
considerations, or the growth dynamics
of snow flakes, or the electrical conduction in carbon nanotubes. 
There is CMP in the field effect transistor, 
modern computer chips and the sensitive SQUID magnetometer that
detects  the feeble electrical activity that goes on in our restless
brain. 

The field is messy but rewarding. Quantization of 
Hall conductance, that won 2 Nobel prizes,
occur amidst disorder and interaction.  
While the field is diverse, there are powerful unifying notions 
and ideas: spontaneous symmetry breaking, order parameter, 
renormalization group, complex collective behavior, quantum coherence, 
chaos etc.  The idea of renormalization group is an an example
that has grown out of the study of condensed matter
systems such as liquid-gas phase transition and Kondo problems - it 
has far reaching application potential including possibility of 
understanding some hierarchical structures in biological systems 
to turbulence in classical fluids.

The field of CMP possesses a deep working knowledge of quantum 
mechanics, both in theory and experiments.  This gives it an 
unique strength and also makes its relation to biology special.  
The stability of atoms, the origin of chemical bonds,
electron transfer, proton tunneling etc. in biology
are truly quantum mechanical.  However, it is fair to say that some 
mysterious leaking of quantum effects to some unexpected aspects 
and domains of biology (such as the origin of consciousness), 
apart from the above obvious ones. are distinct perhaps remote 
possibilities.  CM physicists will not accept such suggestions uncritically, 
but will have a natural edge in unraveling those which turn
out to be meaningful.

Physics gains its predictive power and becomes a quantitative 
science because of the powerful use of mathematics -  
analysis and approximations intertwined with physical
insights, order or magnitude estimates, dimensional analysis and 
many modern mathematical ideas such as homotopy theory,
group theory, algebraic geometry, functions of many complex variables
etc.  In view of the remarkable developments in computers computational 
CMP is becoming very popular and powerful.  Often you can do a 
computer simulation or experiment and create situations that you 
find it hard to create in the laboratory, or study analytically.  

I alluded to the complexity of the study of condensed matter.
This gives it a remarkable ability to suggest new paradigms
through its emergent character\cite{pwa}, that could be helpful 
elsewhere.  The last several decades have seen
some of them: i) spin glass and neural network, and ii) self organized
criticality, power laws etc. These notions may not have solved the 
real problem of biology - but they are some new windows for physicists
to look at this totally new world of biology.  The wealth of phenomena
in condensed matter is sure to provide  seeds of new paradigms 
provided we look for them and and develop a sensitivity to abstract them.  

\section{Nature of Biology as a science}

Like physics, biology is truly an experimental science.  Most of the
problems in biology are far too complex, at the moment, to be analyzed 
threadbare, the way we do in physics with some problems,  using our
existing knowledge and concepts of physics, chemistry and mathematics. 
However, after the revolutionary beginning of the field of molecular 
biology\cite{watson}
at the middle of this century, biology has taken a new shape, and
looks comfortable even
for a physicist to look at from a distance.  Very general principles
like Darwin's natural selection to very specific structure
and function relations in DNA, proteins, etc. are  dominating the field 
currently.  There are also many dogmas, hard earned hypothesis and
working principles that pervade this truly diverse field -
protein structure, signal transduction, brain function to name 
a few.  Thanks to the 
experimental tools like x-ray crystallography, electron microscopy,
NMR imaging, ATM, STM, optical tweezers and so on, that actually
came from physics, the field is undergoing revolutionary development.

The urgent problem facing a hard core biologist is often very different
from what a physicist, genuinely interested in biology, is capable of 
solving in a short time period.  This is the reason many biologists
sincerely feel that physicists can not solve the mysteries of biology.
On the other hand, physicists like Schroedinger, Max Dellbruke, 
Crick, Hopfield and others have made truly original contributions
and opened up new directions.  It is becoming clear that physics 
is not just providing experimental tools to other fields such as 
biology,  it is evolving capabilities and insights to understand 
the spirit of biology.  

\section{Some Examples}

Having made several general remarks let me indicate some examples, 
based on my one decade of a distant admiration for
biology - it is so distant that biology does not know that I am
dreaming of her ! 

{\bf Brain}

A brain would naturally like to think about how it thinks;
why grey matter-a large piece of condensed matter possesses 
consciousness, self awareness, minds eye-I etc.  Physicists
have no clues as to how the laws of quantum mechanics, thermodynamics
or even quantum gravity for that matter, leads to these profound 
properties in a  living state of condensed matter. After listening to 
an illuminating  talk by John Hopfield on neural network and associative 
memory\cite{hopfield} in the fall of 1987 at Princeton, I thought, in a moment of 
weakness,  that I understood the physics of the mind ! - soon to 
realize it is far from it. 

It is becoming increasingly clear that all our understanding  
of spin glass physics, that came from partly the study of a dilute 
concentration of magnetic impurities in an otherwise pure gold, has only 
landed us somewhere at the plains of the Himalayan range, we have
to scale Mount Everest.  The concept of 
network, basin of attraction, possible hierarchical or ultra metric
organization of attractors, are all  but words in constructing a long 
poem.  This became clear to me, when I participated in an Institute
of Theoretical Physics, Santa Barbara workshop
on `Neurobiology for physicists' in the fall of 1987, where 
neurobiologists humbled us with mind boggling biological and clinical
facts about the brain.  But, at the same time, our successful 
understanding of the neural network of sea slug, 
an organism  that has few hundred 
neurons, gives us hope that perhaps one day we can understand 
mammalian brain, with all its complexities and hierarchies.  
It is clear that all our efforts so far has been a warm up exercise.

{\bf Gene}

The next example is from DNA.  I was surprised, like my colleagues, 
by the findings of some physicists\cite{stanley}
that DNA of various living organisms 
possess some kind of long range or power law correlation in its
nucleotide sequence - that is, the probability that two nucleotides
separated by n bases in a strand of DNA, will be both adenine for 
example, is $\approx \frac{1}{n^{\alpha}}$. It was claimed that the
exponent $\alpha$ is species dependent - a one 
parameter characterization of a species at the level of nucleotide
sequence in DNA !  

Soon it became clear that things are much more complicated - there 
are introns, the non coding genes, tandem repeats and so on
that make this long range correlation not very meaningful or
insightful.  However,
it gave an opportunity to many physicists like me, to get a glimpse
of the world of biology with problems that are even more challenging 
than the power law correlation in the nucleotide sequence. 
Genetics is full of many surprises - gene replication, 
gene repair, translation, gene regulation etc.  All my initial 
enthusiasm to model the DNA as a one dimensional  4 state Potts model 
with long range interaction and frustration soon gave way to other 
glamours of biology.

Gene regulation is a fascinating subject.  This is what controls
the shape of a blue whale or a butterfly, in its growth, by
a profound regulation of the production of various proteins, 
at various cell at different times 
starting from the first zygote (formed by the union of a sperm and 
an egg) cell. It is a network that is very different from 
the neural network or immune network or the spin glass or glass.
At the same time it is a network that should posses some general
characteristics of any large networks - this is what prompted
Kaufman, for example, to invent a Boolean net to model gene
regulation\cite{kaufman}.

Physicists have come across some networks and learned some 
general principles - network of dislocation that
controls the mechanical properties of solids under stress, shape 
memory alloys, glasses and spin glasses. Thanks to experiments that 
guides the theoretical developments hand in hand, we have gained some 
insight and useful notions have been developed\cite{network}. 
Erstwhile Condensed 
matter physicists, and my ex collaborator Shoudan Liang is deep into
genetic net and Stan Leibler and Naama Barkai are deep into 
biochemical nets apart from other involvements.  But all our 
insights from CMP are truly warm up exercises at the base camp. 

{\bf Electron and exciton transport in biological systems}

Szent-Gyorgyi, an eminent biochemist, speculated on the importance
of electron transport in biological systems including DNA. He along 
with others have speculated that it could hold some of the the 
secret of carcinogenesis.  Possible connection to cancer apparently
got a lot of (unjustifiable ?) funds  - that is a 
different story.  The point is that the lightest of charged 
particles in biology, namely electron, is involved in too many 
vital activities.
Within proteins electron transport is well studied in biology for
the last many decades.  There are reaction centers, typically a 
prosthetic group such as a porphyrine complexing a metal
ion, embedded in protein.  On absorbing a light quanta the reaction
center releases an electron that tunnels through a couple of tens
of angstrom distance through the folded protein before it is 
absorbed by another special complex, just to trigger another 
reaction; then it continues sometimes ending up in an ion transfer
across the membrane, if it were a membrane bound protein.  

Electron transfer in biological system, even though it takes place
at room temperature, is clearly a quantum mechanical phenomenon.
The theory of Marcus and its generalizations  have been used 
for quantitative estimates of the reaction rates. Our experiences 
with electronic
conduction in semiconductors, metals or in general crystalline 
materials, where Bloch's theory applies is only a warm up exercise
to handle this special disordered system. Condensed matter systems 
such as amorphous materials where Anderson theory of 
localization applies looks too simple and less structured compared
to the mesoscopic biological proteins.  We have a disordered
peptide bond skeleton along with the amino acid side groups
that have considerable number of $\pi$ electrons, where the 
electron correlations are important.  That is there is more structure
including some significant vibrionic couplings and electron correlation
effects.  Most of the present theoretical efforts I have seen are 
one electron type that are computer intensive.  Are we missing some subtle 
effects, including the correlation effect in the $\pi$ electron
pool of the porphyrine rings ? I think only experiments have to 
give an answer to these questions through possible anomalies.
It is interesting that even a diamagnetic response of a pool of 
$\pi$ electrons of planar aromatic ring compounds show  
interesting surprises through correlation effects, a subject 
that worried people like Pauling, London\cite{london}, K S Krishnan and
some in the recent times.  

One hears of new experiments where electron transfer along the 
DNA double helix has been seen\cite{eDNA} indirectly in some experiments.
Its possible relevance in biological functions is an obvious
next question.  Physicists with their warm up exercise and training
in condensed matter can hope to scale the mountain after knowing
many biological details and with help from future experiments.  

Structure and function are catch words in biology often used 
at the level of DNA or enzyme functions. In a different
context, in bacterial photo synthesis certain geometrical arrangement 
of porphyrine complexes have given new insights into mechanism of 
energy transport by exciton. The structure of the basic unit of the so 
called light harvesting complex has been deciphered a couple of 
years ago\cite{exciton1}.
There are two types of ring complexes one containing one concentric
circle of 18 porphyrine molecules that are stacked in a circle like
slides in a circular slide box.  The other contains two concentric 
rings with the 
reaction center complex at the center.  These ring complexes are 
organized on the surface of the cell in some quasi periodic fashion.
The incident photon is absorbed by the porphyrine to create an 
exciton which propagates to end up in the reaction center to activate
an electron transfer reaction.
To this complex geometrical arrangement one can apply, as Ramakrishna
and myself attempted among others\cite{exciton2}, our knowledge of exciton transport
in molecular crystals.  Already there are many surprises - one always
felt that the exciton transport is an incoherent hopping process
at the physiological temperatures. 
But within the ring complex the exciton transport has been shown to 
be coherent experimentally and theoretically.  Our feeling is that 
between the ring complex also , through Forster mechanism,
there is some coherence and possibility of new physics.  

What is remarkable is that the photo synthetic apparatus of
the purple bacteria, is probably the simplest of the lot.  When we 
come to even simple algae and plant leaves, the photo synthetic 
apparatus is much more complex and structured, with light guides
and so on.  In many of these cases we do not know in detail the basic
structural units and their organizations - apart from circular complexes
there are cylindrical complexes and light guides.  What we have learned 
in CMP as exciton transport in semiconductors or 
molecular crystals is truly a warm up exercise when we come to 
this very complex photo synthetic apparatus.  

{\bf Regulated self assembly}

Periodic structures are very dear to condensed matter physicists.
We study how these structures change when we heat a solid or how 
a beautiful sugar crystal grows from a tiny nucleus in a concentrated
sugar solutions.  There is plenty of physics and statistical mechanics.
Some times there are even quantum effects like in the case of Helium 
solids or solids of light elements such as Li.  

In biology very rarely do we come across periodic structures.
Since the structure of proteins and DNA imply important functions,
evolution has not chosen structures that are  manifestly periodic.
However, there are remarkable regular, sometimes symmetric  and 
hierarchies of structures. For example if we look at the virus T4, there are
a few types of  basic proteins that make up the so called pro head -
a complex of proteins that has icosahedral symmetry that encapsulates
the viral DNA.  Then there is a neck, again made up of protein complexes
and a body (that looks like a bit of micro tubule) - a cylinder made up 
of proteins and the legs made of proteins.  This tiny little `robot' 
is different from a periodic crystal.  However there is some regularity
in its making. 

And condensed matter physicist is tempted to wonder about the assembly 
of this complicated macromolecular robot.  The physics is not exactly 
that of the growth of a sugar crystal.  It is
a self assembly that is regulated. It is non equilibrium
statistical mechanics that is embedded in a signaling network.

The regulated self assembly is a
new notion that is very unique to and ubiquitous in biology. 
The above is only an example of the many hierarchical structural 
organizations that one comes across in biology - morphogenesis,
micro tubules, myosin complex, collagen fibers, fibrils etc.  

In fact I learned about this notion of regulated self assembly
during my sabbatical at Princeton in 1996, in a Cell Biology course 
organized by Stan Leibler and Frank Wilczek.  It became clear
in that course, that had distinguished attendees like 
Curtis Callan, Stephen Adler and others, that there are many 
challenging and profound problems.  We physicists returned
spell bound at the end of every class on learning new wonders in 
cell biology and felt the need for serious investigations by many 
physicists.  

Finally a word about some macro molecular structural changes in biology.
Structural rearrangements in biology are in plenty.  A heme protein,
as soon as it gets an oxygen,  undergoes a conformational change
so that it can bind the second oxygen more easily and so on.
An allosteric protein like the  motor protein, once it gets an 
ATP undergoes a massive conformational or structural change, that 
is like an elementary step in walking.  Our well known notions
such as soft modes or anharmonic interactions that we are 
used to in structural changes in simple solids, are far from 
sufficient to understand even the simplest macro molecular structural 
change - we have to think afresh.  People like Frauenfelder, Austin,
Stein, Wolynes and others have made a start at this.  

\section{Conclusion} 
The trend of many condensed matter physicists taking a serious look at 
biology is visible for a long time.  
One also hears of new Institutes and ventures like Santa Fe Institute
which catalyses new kind of activities and exchanges.  In a special
section devoted to `Complexity Science', a recent issue of 
Science\cite{science}
enumerates about a dozen Universities and Labs in the 
United States trying to set up new across the disciplinary ventures
involving physics and biology departments, just at the turn of 
this century.  This has started happening 
in a natural fashion in developed nations like the US or Europe.
The developing countries will do well to recognize this and
participate and contribute to this resonance and redefinition 
among disciplines in science.  

When the condensed matter physics - biology resonance touches the
spirit of biology, the nature of progress will be substantial.

{\bf Acknowledgment}

I thank P.W. Anderson for his critical reading 
of the manuscript and comments and  S. Arunachalam for correcting
the manuscript.


\begin{references}
\bibitem[\ddag]{email}
Baskaran@imsc.ernet.in

\bibitem{pwa} P.W. Anderson, Science {\bf 177} 393 (1972)

\bibitem{watson} see for example, J.D. Watson et al. Molecular 
Biology of the Gene (Benjamin/Cummings Publishing Co. 1987);
B. Alberts et al. Molecular Biology of the Cell (Garland
Publishing Inc., 1994)

\bibitem{hopfield} J. J. Hopfield, Proc. Nat. Acad. Sc. (USA)
{\bf 79} 2554 (82); G. L. Shaw and R. Vasudevan, Mathematical
Biosciences {\bf 12} 207 (1974); W.A. Little, Mathematical 
Biosciences {\bf 12 } 101 (1974)

\bibitem{stanley} C.-K. Peng et al. Nature {\bf 356} 168 (92);
R. Voss, Phys. Rev. Lett. {\bf 68} 3805 (92);
Maria de Sousa Vieiera, cond-mat/9905074

\bibitem{kaufman} S. Kaufman, The Origin of Order (Oxford University
Press, 93)

\bibitem{london} F. London, Phys. Radium {\bf 8} 397 (1937);
see for some modern works, Y. Anusooya and Z.G. Soos, Current 
Science {\bf 75} 1233 (1998)

\bibitem{network} Dynamical Networks in Physics and Biology,
Les Houches School, Ed. D. Beysens and G. Forgacs (Springer 97)

\bibitem{eDNA} Hans Werner Fink and C.Schonenberger, Nature
{\bf 398} 407 (99) 

\bibitem{exciton1} M.Z. Rapiz et al. Trends in Plant Sciences,
{\bf 1} 198 (96) 

\bibitem{exciton2} V.I. Novoderezhkin and A.P. Rajivin,
Chme. Phys. {\bf 211} 201 (96); Ivan Barvik et al., Chem. Phys.
{\bf 194} 117 (95) and a preprint; S. Ramakrishna and G. Baskaran,
unpublished

\bibitem{science} Special section on Complex system Science,
Science {\bf 284} 79 - 107 (1999);  S. Arunachalam, Current
Science {\bf 76} 1191 (99)

\end{references}
\end{document}